\documentclass[aps,prb,twocolumn,showpacs,preprintnumbers,amsmath,amssymb]{revtex4}

\usepackage{epsfig}
\usepackage{dcolumn}
\usepackage{bm}
 
\begin{document}
\title{Thermal transport of 
molecular junctions in the pair tunneling regime}

\author{Karol Izydor Wysoki\'nski} 
 \email{karol@tytan.umcs.lublin.pl}
\affiliation{Institute of Physics,
M. Curie-Sk\l{}odowska University,
Radziszewskiego 10, 
Pl 20-031 Lublin}

\date{ \today}
\begin{abstract}
Charge and heat transport through a single 
molecule tunnel-coupled to external normal electrodes have been studied.
The molecule with sufficiently strong interaction between electrons and 
vibrational internal degrees of freedom  
can be  characterized by the negative effective charging energy U$<0$.
  Such a molecule  has been considered and modeled
  by the Anderson Hamiltonian. 
The electrical conductance, thermopower and thermal conductance of
the system have been calculated as a function of gate voltage
in the weak coupling limit within 
 the rate equation approach. In the linear regime the
 analytic formulae for the transport coefficients in 
 the pair dominated tunneling are presented. 
 The effects found in the nonlinear transport
include {\it inter alia} the rectification  of the heat current. 
The sense of forward  (reverse)
direction, however, depends on the tuning parameter and can be
controlled by the gate voltage.   
We also discuss the quantitation of the thermal conductance 
and the departures from the Wiedemann-Franz law.
\end{abstract}
\pacs{PACS numbers: 73.23.-b;   73.63.Kv; 73.23.Hk}

\maketitle

\section{Introduction}
The appearance of the effective attractive interactions between
electrons in metals \cite{BCS} 
is responsible for the instability of the Fermi surface and the resulting
phenomenon of superconductivity; one of the most spectacular quantum 
effects on a macroscopic scale. The strong local electron -- phonon interaction  
may induce formation of polaron or bipolaron  
quasiparticles which are at the heart of bipolaronic 
theory of superconductivity \cite{micnas1990,alexandrov1994}. 
The so-called negative U centers \cite{anderson1975}   are not only a source 
of superconducting instability but also play
an important role in  physics and chemistry 
of materials \cite{varma1988}  (see, however \cite{harrison2006}). 

The  tendency towards bipolaron formation is enhanced \cite{alexandrov1995}
in the confined structures like quantum dots\cite{jacak1998}. 
This assertion has been corroborated by means of quantum Monte Carlo
calculations in the strong coupling regime for various dimensionalities of
 nano-structures \cite{hohenadler2007}.
Negative charging energy of the small systems
like quantum dots is a relatively new 
concept \cite{Koch,alexandrov2002,holmqvist2008,gierczak2008}. 
Very small system is characterized  by the small capacitance $C$ which makes 
the charging energy $e^2/2C$  large. 
 To overcome it sufficiently strong polaronic shifts resulting  
from strong coupling of vibrational 
and charge degrees of freedom in a molecule are needed. The charge
fluctuations in a device with metallic and superconducting grains
may also over-screen the Kondo repulsion \cite{holmqvist2008} in the normal
grain. 

The purpose of this paper is to extend the recent work \cite{Koch} 
in which charge transport through the molecule \cite{galperin2007}
 characterized by the negative charging energy 
has been studied. Modeling the system using 
Anderson like Hamiltonian with a negative value of the
charging energy $U$ and considering  weak 
coupling between the molecule and the electrodes the authors
have  systematically accounted for the processes contributing 
to charge transport
and calculated its differential conductance.
The study of the negative U molecule has later
been extended to the stronger coupling and the regime of the 
charge Kondo effect\cite{koch2007}.

Here we consider the same model and study the
charge and heat transport {\it via} negative
U molecule in the weak coupling regime when the pair
tunneling processes dominate the transport.  The analytical  
formulas for the thermopower $S$ 
and the thermal conductivity 
$\kappa$ in the linear regime of the low bias voltage $V=(\mu_L-\mu_R)/e$
and the temperature difference $\delta T=T_L-T_R$ have been found. 
$\mu_{L(R)}$ denotes the chemical potential and $T_{L(R)}$ temperature
of the left (right) external metallic bulk electrode.   $e$
denotes the positive electron charge.    
As in the previous work we assume that the molecule 
is characterized by single, doubly degenerated electronic level.
 In the nonlinear regime with large temperature differences
 $\delta T$ and asymmetric coupling to the external leads 
 the system shows heat rectification property. The thermal current 
 and conductance 
 in the forward direction (say, $\delta T>0$) differ from those 
in the reverse direction ($\delta T<0$).

The low temperature thermal conductance $\kappa$ in the linear regime
is proportional to $T$, but the heat quantum, defined as $\kappa/T$ takes 
on a non-universal  value. However, the Wiedemann-Franz law is
obeyed in the limit $T\rightarrow0$. At elevated temperatures 
we observe marked departures from the
Fermi liquid behavior with  the ratio $L= \kappa/ (GT)$ exceeding the
 Lorenz number $L_0=\pi^2k_B^2/3e^2$. 
  
The organization of the rest of the paper is as follows.
In the next section we briefly recall the model
and approach. In section III we  present the transport coefficients  
calculated in the linear regime. In section
IV  the nonlinear transport coefficients are defined and calculated.
In section V we  
discuss the validity of the Wiedemann-Franz  law and quantization of the
 thermal conductance  in the context of pair tunneling. 
We end up with summary and conclusions.

\section{The model and approach}

The aim of this section is to  recall the main results obtained by
Koch {\it et al.} \cite{Koch} and to establish the notation. 
One starts with a general system  consisting of external leads 
and a central molecule coupled   by tunneling  processes. The electron 
subsystem on a molecule interacts 
with  bosonic degrees of freedom (phonons), which are eliminated by
performing the Firsov-Lang canonical transformation. We assume 
 polaronic shifts  large enough to make the effective charging energy 
$U_{eff}=U-2\lambda^2 \hbar\omega$ negative. 
We  denote it simply $U$. 
The motional narrowing of hopping parameters as well as shifts of the dot 
energy level are  absorbed in the definition of the corresponding 
parameters in the  Hamiltonian 

\begin{equation}
H  = H_{mol}+ H_{leads}+H_i\,,
\end{equation}
where 
$ H_{mol}=\varepsilon_dn_d+Un_{d\uparrow}n_{d\downarrow}$ 
describes the interacting electrons on the dot. Here $n_d=\sum_{\sigma} n_{d\sigma}
=\sum_{\sigma}d^{\dagger}_{\sigma} d_{\sigma}$ is 
the number operator, and $d^\dagger_\sigma(d_\sigma)$ denotes 
the creation (annihilation) operator of
a spin $\sigma$ electron on the molecule. 
 The external leads (left-L and right-R) are characterized by  
the  electron energy spectrum $\varepsilon_{\bf k}$. The corresponding 
term in the Hamiltonian is given by
$ H_{leads}=\sum_{\lambda=L,R} \sum_{{\bf k},\sigma} 
\xi_{\lambda, {\bf k}}
c_{\lambda\,{\bf k}\,\sigma}^\dagger
c_{\lambda\,{\bf k}\,\sigma}$, with $\xi_{\lambda\,{\bf k}}
=\varepsilon_{\lambda\,{\bf k}}-\mu_{\lambda}$ and  
 $\mu_{\lambda}=\mu-eV_{\lambda}$ denoting the chemical potential 
of the $\lambda$ electrode subject to the bias voltage $V_{\lambda}$.
In the following we  assume the equilibrium value of
the chemical potential $\mu=0$ and 
$\epsilon_{L{\bf k}} = \epsilon_{R{\bf k}} = \epsilon_{{\bf k}}$.

The coupling between the molecule and leads is governed by the 
Hamiltonian
\begin{equation}
H_{i}=\sum_{\lambda=L,R} \sum_{{\bf k},\sigma} 
(t_\lambda c_{\lambda\,{\bf k}\,\sigma}^\dagger d_\sigma +h.c.)\,.
\end{equation}
To correctly account for all low energy processes which contribute
to the transport in the limit of negative $U$, it is  
advisable to eliminate $H_i$  
by means of the Schrieffer-Wolff \cite{Schrieffer_Wolff}  
transformation. This is valid in the limits $ |t_{\lambda}|\ll |U+\varepsilon_d|$
and $|t_{\lambda}|\ll |\varepsilon_d|$ \cite{schuttler1988}.
 One gets \cite{Koch} the effective
low energy Hamiltonian  
\begin{equation}
\tilde{H}=H_{mol}+H_{leads}+H_{dir,ex}+H_{pair},
\end{equation}
in which the direct and exchange $H_{dir,ex}$
interactions between the dot and the leads read 
\begin{eqnarray} 
&&H_{dir,ex}=\frac{1}{2}\sum_{\lambda \lambda'{\bf k}{\bf k'}\sigma}
{t_{\lambda} t_{\lambda'}^*} \bigg[
\frac{1}{\varepsilon_{\lambda{\bf k}}-\varepsilon_d} 
 c_{\lambda{\bf k}\sigma}^\dagger 
c_{\lambda'{\bf k'}\sigma}  +  \\ \nonumber
&& M(\varepsilon_{\lambda{\bf k}})
 (d_{-\sigma}^\dagger d_\sigma c_{\lambda{\bf k}\sigma}^\dagger c_{\lambda'{\bf k'}\,
-\sigma}
- c_{\lambda{\bf k}\sigma}^\dagger c_{\lambda'{\bf k'}\sigma}n_{d\,\bar{\sigma}})+
H.c.\bigg]\,,\nonumber
\end{eqnarray}
where
$
M(\varepsilon)=[\varepsilon-\varepsilon_d]^{-1}-[\varepsilon-\varepsilon_d-U]^{-1}\,.
$
These terms are most important in the study of 
transport through the quantum dots with positive 
charging energy, but  they do
also play a role in the present case of 'negative $U$ quantum dot' \cite{QD}
and describe single electron co-tunneling processes.

The pair terms read \cite{Koch} 
\begin{equation}
H_{pair}=\sum_{\lambda\lambda'{\bf k}{\bf k'}} t_{\lambda}t_{\lambda'}^* 
M(\varepsilon_{\lambda{\bf k}}) 
d_\uparrow d_\downarrow c_{\lambda'{\bf k'}\downarrow}^\dagger 
c_{\lambda{\bf k}\uparrow}^\dagger+h.c.\,.
\end{equation}

For negative values of $U$, the pair tunneling terms 
 play a main role because the state with two electrons on the
 dot is its lowest energy state and the double occupancy 
 of the dot is favorable. The two electrons may tunnel
 onto the dot from a single lead or from two
 different leads.  The single occupation of the dot is not favored
 for the low voltage bias $eV\ll |U|$ and the sequential
 single particle processes are exponentially suppressed.
 As a result the single particle events can contribute to the
 transport through negative U molecule {\it via} higher order
 processes (co-tunneling).

The approach is based on the rate equations for the
occupation probability $P(n)$, where $n$ denotes a state of
the molecule. Since for negative $U$
the single occupation of the molecule is never favorable,
one finds either one or two electrons on a molecule. The energies 
of these two states are equal if $2\varepsilon_d+U=0$. 
Due to the normalization condition $P(0)+P(2)=1$ 
the rate equations reduce to 
${dP(2)/dt}=P(0)W_{0\rightarrow 2}-P(2)W_{2\rightarrow 0}$.  
In a stationary state this gives $P(0)=W_{2\rightarrow 0}/\left(W_{0\rightarrow 2}
+W_{2\rightarrow 0}\right)$, 
where, in the notation of the paper [\onlinecite{Koch}], 
$W_{i\rightarrow f}$ is the total 
transition rate from the initial state $i$ to the final state $f$. 

To calculate transition rates we use the Fermi's golden 
rule \cite{Bruus}  and consider all processes 
in which electrons or electron  pairs tunnel 
from the electrodes $\lambda$ and $\lambda'$ to
the dot or {\it vice versa}. 
With a constant  temperature   across the system they 
have been obtained by Koch {\it et al.}  and for the 
process $0 \rightarrow 2$ read \cite{Koch}  
\begin{equation}
W^{\lambda\lambda'}_{0\rightarrow 2}=\frac{\Gamma_{\lambda} \Gamma_{\lambda'}}{h} 
\int d\varepsilon M^2(\varepsilon) 
f(\varepsilon-e V_{\lambda})f(2\varepsilon_d+U-\varepsilon-e V_{\lambda'}) 
\label{rate1}
\end{equation}
Equation (\ref{rate1}) describes the transfer rate 
for tunneling of 2 electrons onto the dot (spin up electron hops on the dot 
 from the electrode {$\lambda$} and spin down one from the electrode {$\lambda'$}. 
 In the formula   (\ref{rate1})
  $\Gamma_{\lambda}=2\pi\rho_{\lambda} |t_{\lambda}|^2$ denotes the
  effective coupling between the dot and the lead $\lambda$, and  
  $\rho_{\lambda}$ stands for the density 
of the states at the Fermi level in the electrode $\lambda$. $f_{\lambda}(x)$
is the corresponding Fermi function. Similar expressions can be derived for
all other rates.
The rates of single electron co-tunneling 
from the left to the right electrode, 
without changing the occupancy of the dot $i.e$ for the processes 
$0\rightarrow 0$ and $2\rightarrow 2$, are obtained   \cite{Koch} 
from    $H_{dir,ex}$.
Obviously Eq. (\ref{rate1}) is also valid for system in which 
the left lead is characterized by temperature $T_L$ different from that of
the right lead $T_R$. In this case, however, no general analytical
solution is possible, except in the linear regime.  It is presented in the 
next section.

\section{The linear regime}

First  let us consider the transition rate of an electron from a given state 
in the electrode $\lambda$ into the quantum dot as induced by the co-tunneling 
term in the Hamiltonian. Let the initial state $|0\rangle_d$ of the dot be 
the vacuum $|0\rangle$ 
(no electrons on the dot). 
The initial state of electrodes represents two Fermi surfaces with the 
electrons occupying the states up to $\mu_L(\mu_R)$. 
We denote the initial state of the 
whole system by $|i\rangle$ and its energy by $E_0$. The process in which an 
electron with the spin $\sigma$ from the state ${\bf k}$ in the electrode $\lambda$ 
is transferred to the state
 ${\bf k'}$ in the electrode $\lambda'$ results in the final state 
 $|f\rangle = c_{\lambda k\sigma}c^+_{\lambda'k'\sigma}|i\rangle$. 
 The energy of the final state is $E_f = E_0 + \xi_{\lambda'k'} - 
 \xi_{\lambda k}$.  In the single process the energy 
 transported per unit time from the electrode $\lambda$ is $\xi_{\lambda{\bf k}}$. 
  Its contribution to the total energy flux through the left junction 
  equals 
\begin{eqnarray}
 W^{\epsilon\lambda,\lambda'}_{0\rightarrow 0} &=& 
 {2\pi\over\hbar} \sum_{kk'} |t_\lambda|^2|t_{\lambda'}|^2 
 {\xi_{\lambda {\bf k}}\over 
 (\varepsilon_{\lambda {\bf k}} - \epsilon_d)^2} 
 f_\lambda(\varepsilon_{\lambda {\bf k}}) \\ \nonumber
 & & (1 - f_{\lambda'}(\varepsilon_{\lambda' {\bf k'}})) 
 \delta(\xi_{\lambda {\bf k}} - \xi_{\lambda' {\bf k'}})
\end{eqnarray}
$f_\lambda(\varepsilon_{\lambda {\bf k}}) = 1/(e^{(\epsilon_{\lambda{\bf k}} -
 eV_\lambda)/k_BT_\lambda} + 1)$ is the Fermi distribution function. 

In fact the same  factor $\xi_{\lambda{\bf k}}$ contributes 
to the heat flux in an elementary process in which {\it e.g.} a pair is hopping 
from the states $\lambda{\bf k},\lambda'{\bf k}'$ onto the dot, if $\lambda = L$ 
and $\lambda' = R$ . In this process only the member 
of the pair with the  quantum numbers
 $L{\bf k}$ transports the energy flux through the  left junction. 
 If both electrons stem from the left electrode then their 
 contribution to the total flux reads 
\begin{eqnarray}
 W^{\epsilon\lambda\lambda'}_{0\rightarrow 2} &=& {2\pi\over\hbar}
  \sum_{{\bf k}{\bf k}'} |t_\lambda|^2 |t_{\lambda'}|^2
|M(\varepsilon_{\lambda {\bf k}})|^2(\xi_{\lambda {\bf k}} +
 \xi_{\lambda'{\bf k'}}) \\ \nonumber 
& &  f_\lambda(\varepsilon_{\lambda {\bf k}}) f_{\lambda'}(\varepsilon_{\lambda' {\bf k'}}) \, 
\delta(2\epsilon_d + U - \xi_{\lambda {\bf k}} - \xi_{\lambda'{\bf k'}})\,.
\end{eqnarray}
The conservation of energy expresses the fact that if initially 
the energy of the system is $E_i = E_0$ then final energy is 
$E_f = E_0 + 2\epsilon_d + U - \xi_{\lambda k} - \xi_{\lambda'k'}$. 
In calculations   of the current flux $I_e$ the extra factor 
of 2  appears, accounting for a double charge carried in the above process \cite{Koch}.

The net charge and heat  currents 
(each of them being a sum of pair 
and co-tunneling contributions) in the left junction are given by
\begin{eqnarray}
  I_e &= &-e\left\{P_0\, W^{{\rm tot},L}_{0\rightarrow 2} - 
P_2W^{{\rm tot},L}_{2\rightarrow 0}\right\} \nonumber \\
&&-e\left\{P_0W^{ {\rm tot}}_{0\rightarrow 0} 
+ P_2W^{{\rm tot}}_{2\rightarrow 2}\right\} \nonumber \\
I_Q &= &P_0W^{\epsilon,{\rm tot},L}_{0\rightarrow 2} - 
P_2W^{\epsilon,{\rm tot},L}_{0\rightarrow 2}  \nonumber \\
&&+P_0W^{\epsilon,{\rm tot}}_{0\rightarrow 0} 
+ P_2W^{\epsilon,{\rm tot}}_{2\rightarrow 2}
\end{eqnarray}
with $W^{{\rm tot},L}_{n\rightarrow m} = 2W^{LL}_{n\rightarrow m} 
+ W^{LR}_{n\rightarrow m} + W^{RL}_{n\rightarrow m}$, 
while $W^{{\rm tot}}_{n\rightarrow n} = 
W^{LR}_{n\rightarrow n} - W^{RL}_{n\rightarrow n}$ and 
completely analogous expressions for 
$W^{\epsilon,{\rm tot},L}_{n\rightarrow m}$ and 
$W^{\epsilon,{\rm tot}}_{n\rightarrow n}$. 

In the linear regime $V_{\lambda} \rightarrow 0$, 
$\delta T_{\lambda}\rightarrow 0$ and setting 
$\mu_{\lambda}= \mu +eV_{\lambda}$ we 
express the transition rates up to the linear order in $V$ and $\delta T$, using
\begin{equation}
 f_\lambda(\epsilon) \approx f(\epsilon) + {\partial f\over\partial\epsilon} 
 eV_\lambda -
  {\partial f\over\partial \epsilon} \,\epsilon{\delta T_\lambda\over T}.
 \end{equation}
 For the asymmetric couplings $\Gamma_L \neq \Gamma_R$ the 
 resulting formulae are long and will not be reproduced here. In the special 
 case $\Gamma_L = \Gamma_R = \Gamma$ and for the symmetric distribution of
  voltages $V_{L/R} = \pm V/2$ and the temperature difference 
  $T_{L/R} = T \pm \delta T/2$  we obtain
 \begin{eqnarray}
 I_e & = & G_0 V  + I_e^T \delta T  \nonumber \\
 I_Q & = & I_Q^V V + K \delta T . 
\end{eqnarray}
The contributions to the charge and heat flux 
read
\begin{eqnarray}
 G_0 &=& 2\Gamma_L\Gamma_R {e^2\over h} \bigg[M^2(0) 
 {\beta x\over 2\sinh\beta x} + {f(-x)\over \varepsilon^2_d} \nonumber \\
 &+&  {f(x)\over (\varepsilon_d + U)^2}\bigg]  
	\label{cond} \\
I_e^T&=&-2\Gamma_L\Gamma_R{e\over h} M^2(0) {\beta x^2\over 4\sinh \beta x} {1\over T}	\\
I_Q^V&=&2\Gamma_L\Gamma_R {e\over h} M^2(0) {\beta x^2\over 4\sinh\beta x}  \\
K&=&{2\Gamma_L\Gamma_R\over h} \bigg[\left({f(-x)\over\varepsilon^2_d} + 
{f(x)\over (\varepsilon_d + U)^2}\right)I_2 \nonumber \\ 
& +& M^2(0) {\beta x^3\over 8\sinh\beta x}\bigg]{1\over T}
\end{eqnarray}
In the above formulae we introduced  the notation $x = 2\epsilon_d + U$. 
$x$ measures the distance from the degeneracy point. It can be
tuned by the gate voltage.
Note, that the Onsager reciprocity relation \cite{mahan1981}
 $I_e^T T=-I_Q^V$ is explicitly fulfilled. 

It is  easy to obtain the phenomenological transport coefficients from 
those currents. The linear 
conductance, $G_0$, is defined for $\delta T=0$ as $I_e = G_0V$ 
and is given by equation (\ref{cond}). It has been  discussed earlier
\cite{Koch} and we 
shall not discuss it here.
The thermopower $S$ is defined as the voltage, which appears across the system 
subject to the temperature gradient in the absence of  current flow
\begin{equation}
  S = -\left({V\over\delta T}\right)_{I_e=0}\,.
  \end{equation}
S has been studied earlier \cite{gierczak2008} and the analytic 
formula for it reads  \cite{kiw-expl}
  \begin{equation}
   S = -{k_B\over e} {M^2(0)\beta^2x^2/4\sinh\beta x\over G_0}\,.
  \end{equation}
Thermal conductance $\kappa$ is defined as a coefficients between 
the heat current $I_Q$ and the temperature gradient, under the condition  
of no charge current $I_e=0$. With this definition, we find
\begin{eqnarray}
 \kappa &=& 2\Gamma_L\Gamma_R {k_B\over h}  \bigg[ 
 \left({f(-x)\over \varepsilon^2_d} + 
 {f(x)\over (\varepsilon_d + U)^2} \right) \beta I_2 \nonumber \\ 
 &+& {M^2(0)\beta^2x^3\over 8\sinh\beta x} + 
 S{\beta^2x^2\over 4\sinh\beta x} \bigg]\,.
 \end{eqnarray}
 In the above formula $I_2$, denotes the integral
 \begin{equation}
  I_2 = \int^\infty_{-\infty} 
  d\varepsilon \,\varepsilon^2\left(-{\partial f\over\partial\varepsilon}\right) =
{\pi^2\over 3\beta^2}\,.
   \end{equation}

\begin{figure}[ht]
\begin{center}
\resizebox{0.98\linewidth}{!}{\includegraphics{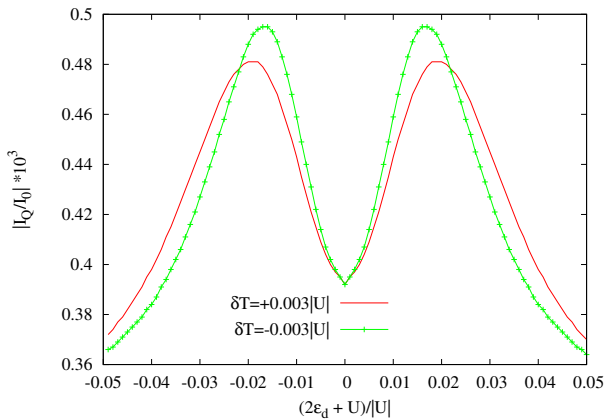}}
\end{center}
 \caption{(Color online) Dependence of the heat current $I_Q$  
 on $x=(2\varepsilon_d +U)/|U|$ for $\delta T=\pm 0.003|U|$ and $\Gamma_L=2, 
  \Gamma_L\cdot\Gamma_R=1$ in units of $|U|$. 
 The temperature $T=0.005|U|$ and $I_0=e |U|/h$.  }
\label{fig-rect}
\end{figure}
  
\begin{figure}[ht]
\begin{center}
\resizebox{0.98\linewidth}{!}{\includegraphics{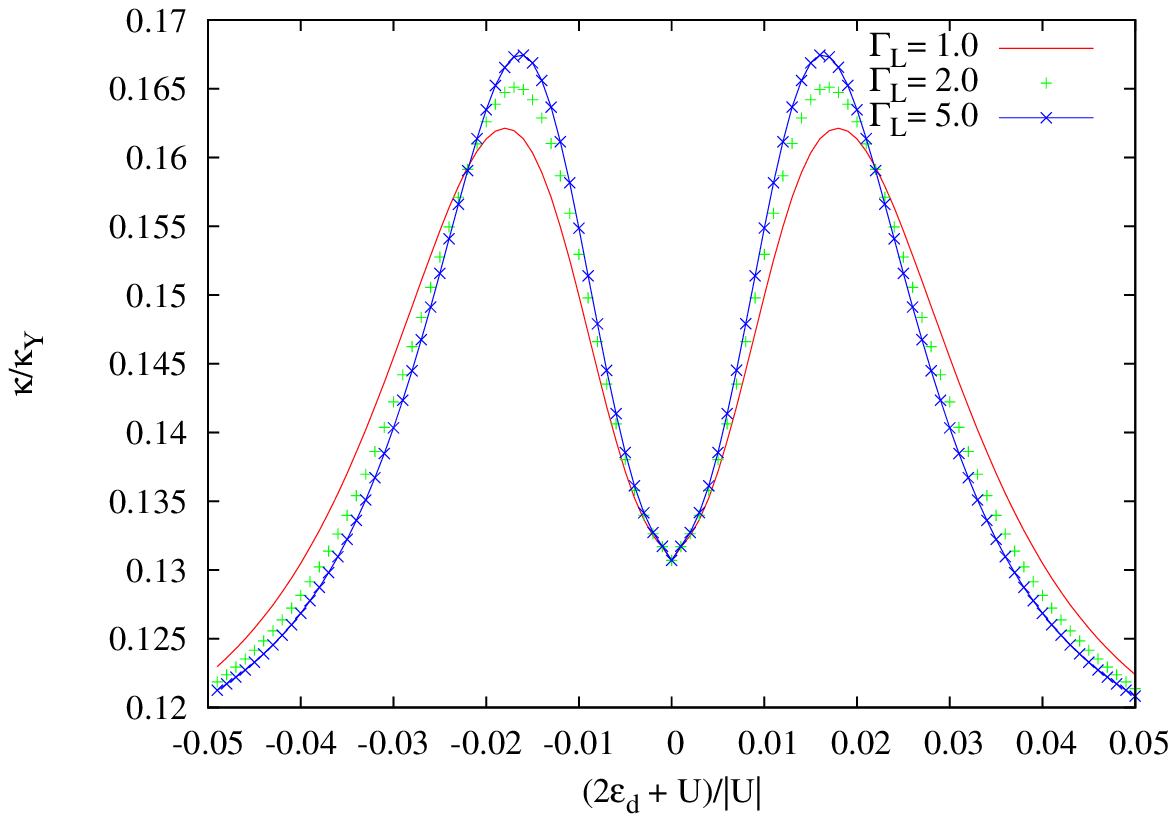}}
\resizebox{0.98\linewidth}{!}{\includegraphics{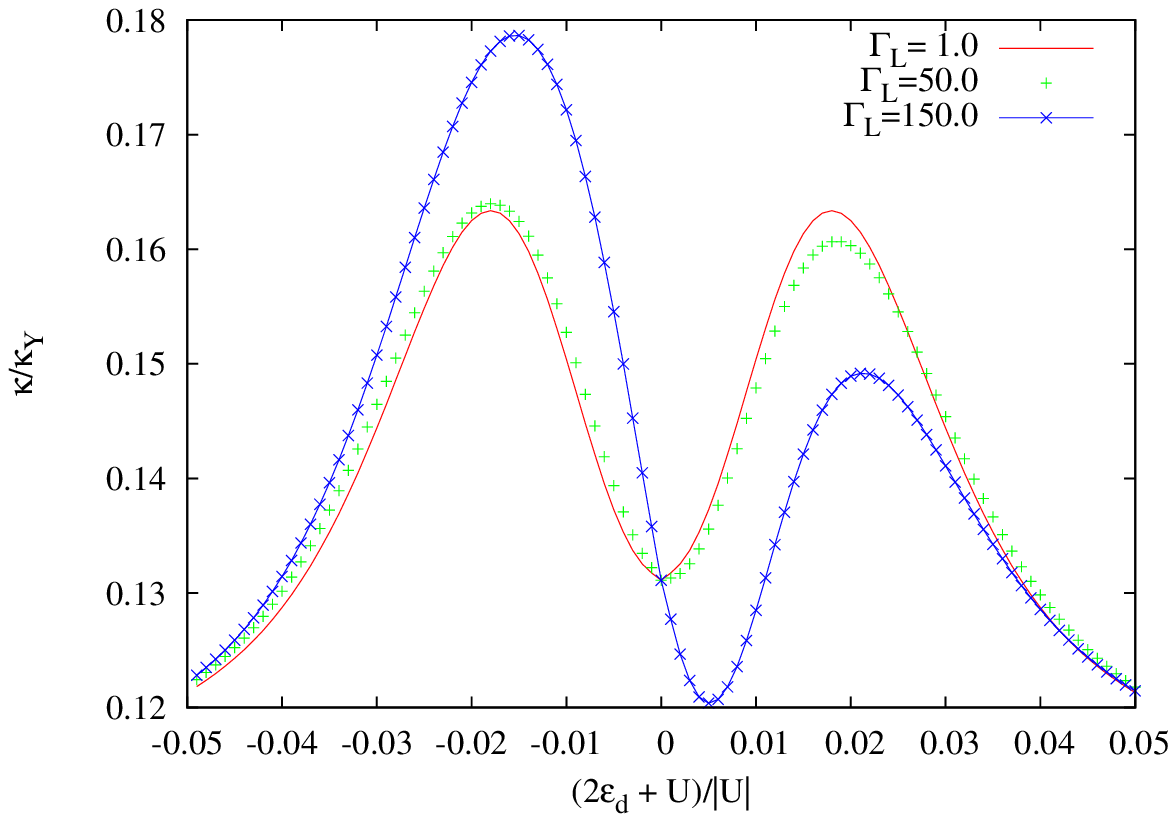}}
\end{center}
 \caption{(Color online) Dependence of the thermal conductance $\kappa$ normalized
 to $\kappa_U=k_B |U|/h$ on 
 $x=(2\varepsilon_d +U)/|U|$ for temperature $T=0.005|U|$, 
  $\delta T=0.001|U|$. Figure a) is for strong junction asymmetry, while b) is
  for very large  asymmetry.}
\label{fig-kappa}
\end{figure}
  
\section{Thermal rectification in a nonlinear transport}
    
Outside the linear regime,  {\it i.e.}   for finite voltage bias $V$ and temperature 
difference between the electrodes, the transfer rates
 $W^{\lambda\lambda'}_{n\rightarrow m}$ and $W^{\epsilon\lambda\lambda'}_{n\rightarrow m}$ 
have to be calculated numerically. In the  studies of charge 
conductance \cite{Koch} interesting rectification properties of the device have 
been predicted. For strongly asymmetric coupling of the dot
 to the electrodes e.g. $\Gamma_L >> \Gamma_R$ the current $I_e$
 is the asymmetric function of $x$ in the limit of $|eV| >> k_BT$. 
The intriguing question arises if similar rectification  properties \cite{thermal-rect}
could be achieved in the heat flow. To answer this we have 
calculated the currents 
    $I_e$ and $I_Q$ for general values of $V$ and $\delta T$. 
Formally both charge and heat currents depend on a voltage
and a temperature difference 
    \begin{eqnarray}
    && I_e = I_e(V, \delta T) \\
    && I_Q = I_Q(V, \delta T)\, .
    \label{heatcur} 
    \end{eqnarray}

 As stated earlier the thermal conductance $\kappa$ is 
generally defined by 
$I_Q = -\kappa\delta T$ under the condition $I_e = 0$. The condition of no 
current flow   defines the
 thermoelectric 
power $S = -\left({V\over\delta T}\right)_{I_e=0}$, where $V$ is the voltage
generated in the system by the temperature gradient.
It provides the 'selconsistent value' of the voltage 
to be used in equation (\ref{heatcur}). 
Introducing this into the equation for $I_Q$ we get
\begin{equation}
 I_Q =I_Q(V=-S\delta T,\delta T)=-\kappa \delta T\,.
 \label{nlthermal-cond}
 \end{equation}
The last equality is our definition of the thermal conductance $\kappa$.
In this section we measure all energies in units of $|U|$. 
The dependence of $\kappa$ on $x=2\varepsilon_d +U$ is 
shown in  Fig. (\ref{fig-kappa}) for the temperature $T=0.005|U|$
and relatively small asymmetry of the junction, parametrised by
the value of $\Gamma_L$ in the units of $|U|$, with $\Gamma_L\cdot\Gamma_R=1$ 
in the same units.

Calculating the currents we have assumed symmetric voltage and 
temperature 
difference with $V_{L,R}=\pm V/2$ and $T_{L/R}=T \pm \delta T /2$. 
For positive $\delta T$ the heat will
normally flow from the left to the right lead. Thermal rectification can be 
defined \cite{thermal-rect} as the dependence of the heat current $|I_Q|$
or the nonlinear thermal conductance $\kappa$ defined in Eq. (\ref{nlthermal-cond})  
on the sign of temperature  difference.
Figure (\ref{fig-rect}) shows the effect of heat rectification 
in the asymmetric molecular junction 
with $\Gamma_L/\Gamma_R=4$, the temperature $T=0.005|U|$
and two values of $\delta T =\pm 0.003|U|$.  
 Due to phase space restrictions, which make the transition rates 
 for two -- electron processes $V$ and $\delta T$ dependent, 
 the excess heat current $\delta I_Q=|I_Q(+\delta
T)|-|I_Q(-\delta T)|$ depends on $x$ and changes sign 
around $|x| \approx 0.022|U|$ for a given set of parameters. 

Both the heat flux and  thermal conductance are the symmetric 
functions of $x$ for small
values of asymmetry (see Fig. (\ref{fig-kappa}a)). However, 
for very strongly asymmetric junctions, the
conductance starts to be the asymmetric function of the detuning parameter. 
This is shown in the Fig. (\ref{fig-kappa}b).
The rectification coefficient defined as 
$\delta I_Q/I_Q$, changes sign as the function of $x$. For the 
parameters in figure (\ref{fig-kappa}a) it  
takes the maximal value of about 5\%. 

\begin{figure}[ht]
\begin{center}
\resizebox{0.98\linewidth}{!}{\includegraphics{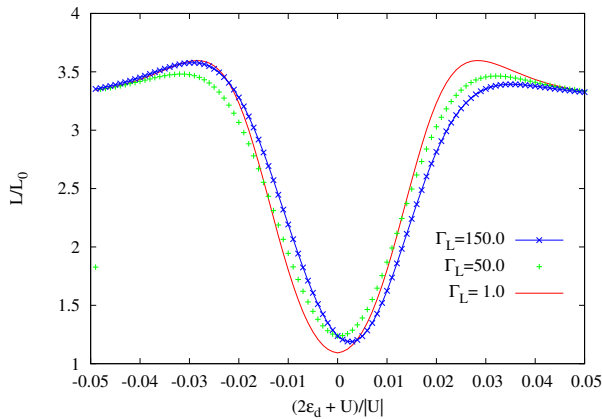}}
\end{center}
 \caption{(Color online) Dependence of the Wiedemann-Franz ratio L 
 normalized  to its Fermi liquid value $L_0={\pi^2 k^2_B \over 3e^2}$ 
on the parameter $x=2\varepsilon_d +U$  in units of $|U|$ for $T=0.005|U|$
and $\delta T=0.001|U|$. Couplings to the leads measured in  units of $|U|$
are normalized by $\Gamma_L \cdot \Gamma_R=1$. }
\label{fig-WFlin}
\end{figure}

\section{Wiedemann-Franz ratio and quantization of electron thermal conductance}

   It has been predicted \cite{rego1998}  that at low temperature the phononic 
thermal conductance of one dimensional dielectric wire is universally given by
   \begin{equation}
    \kappa_0 = \left({\pi^2\over 3}\right) {k^2_B\over h} T
    \end{equation}
leading to the universal, material and temperature independent 
ratio $\kappa_0/T = {\pi^2\over 3} {k^2_B\over h}$, which sets
a fundamental quantum limit on heat flow \cite{schwab2000} in an 
analogy to quantized charge conductance.  
The quantized value of the heat conductance is expected
\cite{rego1998}  in one-dimensional structures  independently of 
the statistics of heat carriers. So this is valid for phonons, 
electrons and also particles with fractional 
statistics \cite{haldane 1991}. This result 
has  been   experimentally  \cite{schwab2000} confirmed 
for phonons, electrons and photons. 

It is an easy exercise 
to show that at $T \rightarrow 0$ the thermal conductance 
through our negative $U$ molecule reduces to
   \begin{equation}
    \kappa(T\rightarrow 0) = 2{\Gamma_L\over |U|} {\Gamma_R\over |U|} \, 
    {\pi^2\over 3} \, {k^2_B\over h} T \,
     {4\over (1 + |x|)^2}\,.
     \end{equation}
  Thermal conductance, which in 
the present context consists of only electron contribution 
is linear in temperature at low temperatures. The heat quantum,
 however, is not universal and depends on the coupling amplitudes
$\Gamma_{L/R}$ 
and the dimensionless distance $x$ from the 
degeneracy point $2\epsilon_d + U = 0$, measured in 
units of $|U|$. It is important to notice that the 
linear charge conductance at very low temperatures
takes on the $T$ independent value
      \begin{equation}
       G_0(T=0) = 2{\Gamma_L\over |U|} {\Gamma_R\over |U|}
 {4\over (1 + |x|)^2} {e^2\over h}\,,
       \label{condT0}
       \end{equation}
    which ensures the validity of the Wiedemann-Franz (WF) ratio in this limit
    \begin {equation}
      {\kappa(T\rightarrow 0) \over G(T\rightarrow 0)T} = {\pi^2 k^2_B\over 3e^2} = L_0\,,
    \end{equation}
    being one of the signatures of the Fermi liquid. 
 
     For an arbitrary  temperature and in the non-linear
    regime the above ratio of thermal to charge
    conductance takes on $T$, $\delta T$ and $x$
 dependent values $L(x,T)={\kappa \over  G T}$. 
 The function $L(x,T)$ in units of the Lorenz number 
  $L_0$ is
 shown   in  figure (\ref{fig-WFlin}) for $T=0.005|U$ and 
 $\delta T=0.001|U|$. At low   $x$ it takes values close to 1, 
 but for larger $x$  one observes 
strong departures  from the Fermi liquid value $L_0$, like in the 
quantum dots with a large Coulomb interaction \cite{kubala 2008}.
Its dependence on the coupling asymmetry is rather weak.

\begin{figure}[ht]
\begin{center}
\resizebox{0.98\linewidth}{!}{\includegraphics{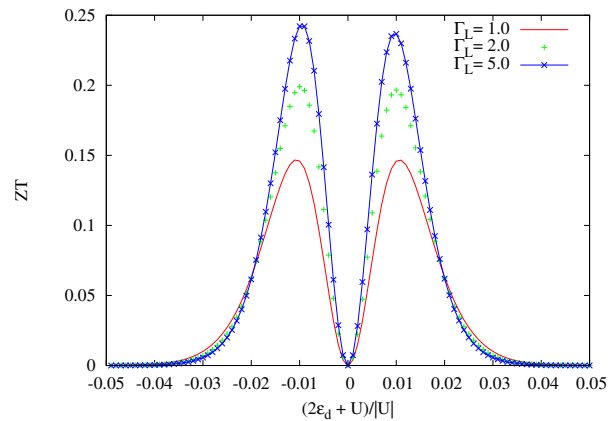}}
\end{center}
 \caption{(Color online) Dependence of the nonlinear dimensionless figure
 of merit $ZT$  on the parameter $x=(2\varepsilon_d +U)/|U|$. 
 The temperature $T=0.005|U|$ and $\delta T=-0.003|U|$ and 
 the couplings are normalized
 by $\Gamma_L\cdot\Gamma_R=1$ in units of $|U|$.}
\label{fig-ZT}
\end{figure}
   
We have also calculated the dimensionless figure of merit which
indicates heat to voltage conversion efficiency of the device \cite{heremans2004}.
It is usually defined as $ZT= G S^2T/\kappa$. For most
devices thermal conductance 
contains both electron and phonon contributions. In the present geometry
phonons do not contribute. The dependence of $ZT$ on $x$ for a few values of
asymmetry is shown in  Fig. (\ref{fig-ZT}). Its value 
increases with the asymmetry of the coupling and is a non-monotonous 
function of $x$. The maximum increases with the anisotropy of 
coupling as it is evident in the figure.

\section{Summary}

We have calculated the linear  thermoelectric transport coefficients  
of the single
electron molecular transistor in the limit of large
 electron-phonon coupling leading to a negative effective 
charging energy. In this limit   the pair tunneling processes  
contribute mainly  to charge and heat transport. 
Pair dominated transport shows strong departures from the 
Fermi liquid characteristics. We have found strong deviations from the
Wiedemann-Franz law  and a non-universal value 
 of thermal conductance  quantum. The thermal quantum {\it i.e.}
 the ratio $\kappa/T$ evaluated at a low temperature and the Lorentz
 function $L(x,T)$ depend on $x$ and asymmetry of the couplings.  
 The thermoelectric figure of merit $ZT$  characterizes 
efficiency of the device for thermoelectric
 applications. This parameter takes on fairly low, albeit 
 in general $x$ and $T$ dependent values for the negative $U$ 
 quantum dot studied here. The maximal value of $ZT$ is about 0.25 for 
 the strongly asymmetric junction.

{\bf Acknowledgmements:} This work has been partially supported 
by the  Ministry of Science and Education under 
the grant No. N202 1878 33 and the scientific network
LFPPI.


\begin{thebibliography}{99}
\bibitem{BCS} J. Bardeen, L. Cooper, J. Schrieffer, Phys. Rev. {\bf 108}, 1175 (1957).
%
\bibitem{micnas1990} R. Micnas, J. Ranninger and S. Robaszkiewicz  
Rev. Mod. Phys  {\bf 62} 113 (1990).

\bibitem{alexandrov1994} A.S Alexandrov, and N.F. Mott, Rep. Prog. Phys. 
{\bf 57} 1197 (1994).
%

\bibitem{anderson1975} P. W. Anderson, Phys. Rev. Lett. 34, 953 (1975);
R. A. Street and N. F. Mott, Phys. Rev. Lett. 35, 1293 (1975).

\bibitem{varma1988} C.M. Varma, Phys. Rev. Lett. {\bf 61}, 2713 (1988).

\bibitem{harrison2006} W. A. Harrison, Phys. Rev. B {\bf 74}, 245128 (2006).

%
\bibitem{alexandrov1995} A. S. Alexandrov and N. F. Mott, 
{\it Polarons and Bipolarons} World Scientific, Singapore, 1995.

\bibitem{jacak1998} L. Jacak, P. Hawrylak  and A. W\'ojs, 
 {\it Quantum Dots} (New York: Springer) 1998.

\bibitem{hohenadler2007} M. Hohenadler and P.B. Littlewood, Phys. Rev. B
{\bf 76} 155122 (2007); M. Hohenadler and H. Fehske, J. Phys.: Cond. Matt.
{\bf 19} 255210 (2007).

\bibitem{Koch} J. Koch, M.E. Raikh  and F von Oppen   Phys. Rev. 
Lett. {\bf 96} 056803  (2006).


\bibitem{alexandrov2002} A. S. Alexandrov, A. M. Bratkovsky, 
and P. E. Kornilovitch, Phys. Rev. B {\bf 65}, 155209 (2002).


\bibitem{holmqvist2008} C. Holmqvist, D. Feinberg, and A. Zazunov,
Phys. Rev. B {\bf 77}, 054517 (2008).


\bibitem{gierczak2008} M. Gierczak, and K.I. Wysoki\'nski, 
J. Phys. Conf. Series, {\bf 104} 12005 (2008).

\bibitem{galperin2007} M. Galperin, M. A. Ratner and A. Nitzan, 
J. Phys. Cond. Matt.  {\bf 19}, 103201 (2007). 

\bibitem{koch2007} J. Koch, E. Sela, Y. Oreg, and F. von Oppen,
Phys. Rev. B {\bf 75}, 195402 (2007).


\bibitem{Schrieffer_Wolff} J. R. Schrieffer and P.A. Wolff  
   Phys. Rev.  {\bf 149}, 491 (1966).

\bibitem{schuttler1988} H.-B. Sch\"uttler and A. J. Fedro, Phys. Rev. B {\bf 38} 9063 (1988).   
   
\bibitem{QD} We use  the notion quantum dot (QD) in a loose
sense to describe small central region of the 
considered structure, independently if it is defined
in two dimensional electron gas, consists of a metallic or semiconducting 
grain or is in the form of a single molecule.


\bibitem{Bruus} H. Bruus  and K Flensberg  {\it Many-body quantum theory
in condensed matter physics}, Oxford Graduate Texts (New York: 
Oxford University Press) (2004), Ch. 10  

\bibitem{mahan1981} G. D. Mahan, {\it Many-Particle Physics}, Plenum Press, 
New York and London (1981), Ch. 3.8.

\bibitem{kiw-expl} The formula for linear thermopower 
derived previously \cite{gierczak2008} is expressed
in terms of the integrals. Unfortunetly, the figures 
presented in that paper 
contain  numerical error which lead to small departures 
(around 2\%  at the extrema of $S$) from the exact result.   

\bibitem{thermal-rect}  G. Casati, Chaos {\bf 15}, 015120 (2005);
 B. W. Li, L. Wang, G. Casati, Phys. Rev. Lett. {\bf 93}, 184301
(2004);  D. Segal, A. Nitzan, Phys. Rev. Lett. {\bf 94}, 034301
(2005);  M. Terraneo, M. Peyrard, G. Casati, Phys. Rev. Lett. {\bf 88},
094302 (2002); C.W. Chang, D. Okawa, A. Majumdar, and A. Zettl, Science {\bf 314},
1121 (2006);  C. R. Otey, W.T. Lau, S. H. Fan, Phys. Rev. Lett. {\bf 104} 154301 (2010);
L. Wang and B. Li, Phys. World {\bf 21}, 27 (2008);
Chen XO, Dong B, Lei XL, Chin. Phys. Lett.  {\bf 25 }, 3032 (2008).

\bibitem{rego1998} J. B. Pendry, J. Phys. A: Math. Gen. {\bf 16} 2161 (1983);
 L. G. C. Rego and G. Kirczenow, Phys. Rev. 
Lett. {\bf 81}, 232 (1998); Phys. Rev. B {\bf 59}, 13080 (1999)

\bibitem{haldane 1991} F. D. M. Haldane, Phys. Rev. Lett. {\bf 67}, 937 (1991)

\bibitem{schwab2000} K. Schwab, E. A. Henriksen, 
J. M. Worlock, M. L. Roukes, Nature {\bf 404}, 974 (2000);
O. Chiatti, J.T. Nicholls, Y.Y. Proskuryakov, N. Lumpkin, I. Farrer, and
D. A. Ritchie, Phys. Rev. Lett. {\bf 97}, 056601 (2006);
M. Meshke, W. Guichard and J. P. Pekola, Nature {\bf 444}, 187 (2006). 

\bibitem{kubala 2008} B. Kubala, J. K\"onig, and J. Pekola, Phys. Rev. Lett. 
{\bf 100}, 0066801 (2008).

\bibitem{heremans2004} J.P. Heremans, C.M. Thrush, and D.T. Morelli,
Phys. Rev. B {\bf 70} 115334 (2004); J.P. Heremans, Acta Physica Polonica
{\bf 108}, 609  (2005); M. Krawiec and K.I. Wysoki\'nski, 
  Phys. Rev. B {\bf 73}, 075307 (2006);
M. S. Dresselhaus, G. Chen, M. Y. Tang, R. G. Yang, H. Lee, D. Z.
Wang,  Z. F. Ren, J.-P. Fleurial, and P. Gogna,
Adv. Mat. {\bf 19} 1043 (2007).

\end{thebibliography}
\end{document}